\documentclass[journal]{IEEEtran}

\usepackage{cite}
\usepackage{url}
\usepackage{amsmath,amssymb,amsfonts}
\usepackage{algorithmic}
\usepackage[utf8]{inputenc}
\usepackage{graphicx}
\usepackage{hyperref}
\usepackage{textcomp}
\usepackage{multirow}
\usepackage{xcolor}

\def\BibTeX{{\rm B\kern-.05em
    {\sc i\kern-.025em b}\kern-.08em
    T\kern-.1667em\lower.7ex\hbox{E}\kern-.125emX}}

\begin{document}

\title{Lightweight Physics-Aware Zero-Shot Ultrasound Plane-Wave Denoising}

\author{
Hojat Asgariandehkordi*,
Mostafa Sharifzadeh*,
Morteza Rezanejad,
and Hassan Rivaz
\thanks{
This work was supported by the Natural Sciences and Engineering Research Council of Canada (NSERC) and Fonds de recherche du Québec (FRQNT).
}
\thanks{
Hojat Asgariandehkordi, Mostafa Sharifzadeh, Morteza Rezanejad, and Hassan Rivaz are with the Department of Electrical and Computer Engineering, Concordia University, Montreal, QC H3G 1M8, Canada 
(e-mail: Hojat.Asgariandehkordi@mail.concordia.ca; 
mostafa.sharifzadeh@mail.concordia.ca; 
morteza.rezanejad@concordia.ca; 
hrivaz@ece.concordia.ca).
}
\thanks{*The authors contributed equally to this work.}
}

\maketitle

\begin{abstract}

Ultrasound Coherent Plane-Wave Compounding (CPWC) enhances image contrast by combining echoes from multiple steered transmissions. While increasing the number of steering angles generally improves image quality, it significantly reduces frame rate and may introduce blurring artifacts in fast-moving targets. In addition, compounded images remain susceptible to noise, particularly when acquired using a limited number of transmissions. In this work, we propose a lightweight physics-aware zero-shot denoising framework for low-angle CPWC ultrasound imaging that improves image quality without requiring external training datasets or clean reference images. The proposed approach partitions the available steering angles into two disjoint subsets, each used to reconstruct compounded images with different angle-dependent artifacts and noise characteristics. These reconstructed images are then used as pseudo-pairs within a self-supervised residual learning framework to train a lightweight convolutional neural network directly on the test sample. Because the underlying tissue structures remain consistent across the subsets while the incoherent artifacts vary with steering angle selection, the proposed physics-aware pairing strategy enables the network to distinguish anatomical information from inconsistent noise and artifacts. Unlike supervised approaches, the proposed method does not require domain-specific fine-tuning or paired datasets, making it adaptable across different anatomical regions and acquisition settings. Furthermore, the proposed framework employs an efficient architecture composed of only two convolutional layers, enabling fast and computationally inexpensive training. Experimental evaluations on simulation, phantom, and \textit{in vivo} ultrasound datasets demonstrate that the proposed method achieves improved contrast enhancement and structural preservation compared to both classical denoising techniques and deep learning-based baselines.

\end{abstract}


\section{Introduction}
\label{sec:introduction}

Ultrasound imaging is a non-invasive and cost-effective diagnostic modality widely used in clinical practice due to its real-time imaging capability, portability, and safety. Despite these advantages, ultrasound images are commonly affected by speckle noise and acquisition artifacts caused by wave interactions with heterogeneous biological tissues~\cite{10423849}. These degradations can reduce image interpretability and diagnostic reliability, motivating the development of advanced image enhancement and denoising techniques~\cite{Liu2025,ESLAMI2022106808}.

Early studies in ultrasound image enhancement mainly focused on classical filtering and signal processing approaches, including spatial filtering, frequency-domain filtering, speckle reduction methods, and contrast enhancement algorithms. Although these techniques can improve visual quality to some extent, they often struggle to preserve fine anatomical structures while effectively suppressing noise~\cite{8808885}.

More recently, deep learning-based approaches have demonstrated remarkable success in medical image restoration tasks, including ultrasound denoising. In particular, supervised convolutional neural networks (CNNs) have been widely employed for speckle suppression, beamforming enhancement, and image restoration~\cite{Luchies2018,Goudarzi2020,Liu2025,Xiao2025}. By learning mappings between noisy and clean image domains, these models can substantially improve image quality.

However, supervised approaches rely heavily on large-scale paired datasets containing noisy and clean reference images. In ultrasound imaging, obtaining such ground truth data is extremely challenging because truly noise-free acquisitions are generally unavailable, especially in clinical patient data. Consequently, many supervised studies rely on synthetic simulations or phantom datasets, which may not generalize well to real-world imaging scenarios. In addition, supervised models are often sensitive to domain shifts caused by variations in transducer hardware, acquisition protocols, or anatomical regions.

To reduce dependence on clean labels, self-supervised denoising methods have been introduced. Techniques such as Noise2Void, Noise2Self, and Neighbor2Neighbor learn denoising functions directly from noisy observations by exploiting spatial redundancy or masked prediction strategies~\cite{Krull2019,Batson2019,Huang2021}. These approaches have also been adapted to ultrasound imaging and other medical modalities~\cite{10584319,Goudarzi2023,Cho2024,Yu2025}. Although self-supervised methods eliminate the need for paired ground truth data, they still require large and diverse training datasets to achieve robust performance. Furthermore, their generalization ability remains limited when applied to unseen domains or acquisition conditions.

In contrast, zero-shot denoising methods operate directly on the test image without requiring external training datasets. One representative example is Zero-Shot Noise2Noise (ZS-N2N), which trains a lightweight network using internally generated noisy image pairs obtained through spatial downsampling~\cite{Mansour2023}. While such methods have shown promising results in natural image processing, their application to ultrasound imaging remains limited, particularly in the context of Coherent Plane-Wave Compounding (CPWC).

In CPWC imaging, multiple plane-wave transmissions acquired at different steering angles are coherently combined to improve image quality and contrast. However, increasing the number of steering angles reduces imaging frame rate and may introduce motion-related artifacts in dynamic imaging scenarios. Moreover, low-angle CPWC acquisitions remain vulnerable to incoherent noise and artifacts.

In this work, we propose a lightweight physics-aware zero-shot denoising framework specifically designed for low-angle CPWC ultrasound imaging. Instead of relying on spatial downsampling, the proposed method leverages the physical properties of plane-wave ultrasound acquisition. Specifically, the available steering angles are divided into two complementary subsets, each used to reconstruct a compounded image. Although both reconstructed images share the same underlying anatomical structures, they contain different angle-dependent artifacts and incoherent noise components. These two images are then used as pseudo-pairs within a self-supervised residual learning framework.

The proposed lightweight CNN is trained directly on the test sample to estimate inconsistent artifacts and noise. After training, the estimated residual noise is subtracted from the original compounded image to obtain the final denoised output. Since the framework does not require any external training data or domain-specific fine-tuning, it can be flexibly applied across different anatomical regions and imaging configurations.

The main contributions of this work are summarized as follows:

\begin{itemize}

\item We introduce a zero-shot denoising framework for CPWC ultrasound imaging that does not require external training data or clean reference images.

\item We propose a physics-aware angular partitioning strategy that exploits angle-dependent artifact variations while preserving consistent tissue structures.

\item We design a lightweight CNN architecture consisting of only two convolutional layers, enabling efficient and computationally inexpensive training.

\item We perform extensive evaluations on simulation, phantom, and \textit{in vivo} datasets, demonstrating improved contrast enhancement and structural preservation compared to classical and deep learning-based denoising approaches.

\end{itemize}

\section{Related Work}
\label{sec:related_work}

A wide range of methods has been proposed for ultrasound image denoising, including classical filtering approaches, supervised deep learning models, and self-supervised learning techniques. In the following section, we review the most relevant literature in these categories.

\section{Related Work}\label{sec:related_work}
A wide range of techniques has been developed to address the denoising problem in ultrasound imaging, ranging from traditional signal processing filters to modern deep learning and self-supervised methods. In this section, we provide a review of related work across three main categories: classical denoising techniques, supervised deep learning-based models, and self-supervised learning strategies.

\subsection{Classical Denoising Methods}

Ultrasound imaging inherently suffers from speckle noise and artifacts due to acoustic wave interactions with tissue structures. Traditional denoising methods including spatial filters, anisotropic diffusion, wavelet transforms, and Non Local Means (NLM) have been widely used to suppress such degradations by leveraging statistical and spatial redundancies~\cite{Coupe2009}. Among them, NLM is notable for preserving anatomical details through patch-based averaging, but it remains computationally intensive and sensitive to parameter tuning. Similarly, BM3D-based approaches, such as the brushlet-enhanced method by Gan \textit{et al.}~\cite{Gan2015}, improve noise suppression but exhibit comparable sensitivity to parameters and computational complexity.

Beam steering has been used to enhance the quality of elastograhy~\cite{Montaldo2009} and speckle characterization~\cite{Rivaz2007}. Advanced classical methods leverage signal coherence in steered angles to improve ultrasound imaging. In CPWC~\cite{Montaldo2009}, post-beamforming coherence-based filters such as the coherence factor~\cite{Hollman1999}, generalized coherence factor~\cite{Li2003}, and their statistical variants~\cite{Yang2020,Wang2019} enhance image quality by promoting phase-aligned signals and suppressing incoherent components. These methods are particularly effective in reducing sidelobe artifacts in beamformed images.

Despite these advancements, classical models remain limited by their heuristic nature, parameter sensitivity, lack of data adaptivity, and computational complexity. Many require careful tuning for specific imaging setups and often underperform in diverse or complex scenarios. Additionally, they often degrade important structural information. These limitations have motivated the rise of deep learning approaches, which offer greater flexibility and higher-quality denoising with significantly reduced inference cost.

\subsection{Supervised Deep Learning-based Denoising Methods}

Deep learning has become foundational in advancing plane-wave ultrasound imaging, particularly in tasks such as beamforming, denoising, and reconstruction from sparsely acquired data~\cite{10584319, Asgariandehkordi2024,Yang2024}.

Early supervised architectures such as the multichannel, multiscale CNN proposed by Zhou \textit{et al.} successfully reconstructed high spatial–temporal resolution plane-wave images by combining multi-angle data and cascading wavelet filters~\cite{Zhou2018}. Nguon \textit{et al.} extended this line of work with a modified U-Net beamformer adapted to multiple imaging conditions through transfer learning~\cite{Nguon2022}.  Goudarzi and Rivaz demonstrated that CNNs trained on simulated (Radio Frequency) RF data generalized effectively to \textit{in vivo} images~\cite{Goudarzi2022}. A CNN beamformer combining GoogLeNet and U-Net was proposed in~\cite{Lu2022} to reconstruct a high-quality ultrasound image using a single-angle plane-wave input. Perdios \textit{et al.} \cite{Perdios2018} trained an adapted version of U-Net on simulation data and showed that their network also performs well on \textit{in vivo} data. Gupta \textit{et al.} \cite{Gupta} looked at image denoising as an inverse problem and solved it using a deep learning model. A network called Mimicknet \cite{Huang} was proposed to improve image quality in post-beamformed data. Van Sloun \textit{et al.}~\cite{8808885} have explored deep learning methods that have been developed for adaptive beamforming, spectral doppler, clutter suppression, and super-resolution.

Compared to natural images, ultrasound data present unique difficulties due to their low contrast and broad dynamic range~\cite{8990076}. These characteristics complicate the direct application of CNNs for image enhancement across various ultrasound tasks. To overcome these challenges, researchers have explored a combination of pre-processing strategies and network designs. The Mean Signed Logarithmic Absolute Error (MSLAE) was introduced as a specialized loss function within a residual CNN framework incorporating multiscale, multichannel filters~\cite{Perdios}. For high-frequency signal recovery, wavelet-based multi-resolution networks have also been proposed~\cite{Goudarzi2022}. Additionally, to avoid convergence to poor local optima in the early training stages, caused by rapid RF fluctuations, a hybrid loss function was designed that adaptively shifts from B-mode loss to RF-domain loss throughout the training~\cite{10584319}. 

Generative Adversarial Networks (GANs) have further elevated perceptual image quality. Zhou \textit{et al.} introduced an ultrasound-transfer GAN to approximate line-scan quality from plane-wave inputs~\cite{Zhou2020}, Tang \textit{et al.} combined adversarial training with attention mechanisms to recover fine anatomical details~\cite{Tang2021}, and Goudarzi \textit{et al.} applied GANs for multi-focus reconstruction~\cite{Goudarzi2020J}. 

Recent attention has shifted toward probabilistic generative models known as Denoising Diffusion Probabilistic Models (DDPMs)~\cite{Ho2020} represent a new paradigm in learning the distributions. DDPMs use a noise reversal process to recover high-fidelity outputs. Enhanced DDPMs have addressed issues like slow sampling~\cite{Nichol2021}, limited expressivity~\cite{Xiao2022}, and high memory cost~\cite{Chung2022},~\cite{Liu2022}.

In ultrasound, DDPMs have been leveraged for denoising and super-resolution. Asgariandehkordi \textit{et al.} trained a DDPM on CPWC data with angular diversity to isolate incoherent noise and recover clean structure~\cite{Asgariandehkordi2024}. Liu \textit{et al.} proposed a diffusion-based super-resolution pipeline for ultrasound~\cite{TianLiu2024}, while Zhang \textit{et al.} demonstrated denoising from under-sampled plane-wave inputs using learned priors~\cite{Zhang2023diffusion}. In \cite{10423849}, a diffusion model was employed for ultrasound cardiography dehazing. Two diffusion models were trained to generate a patch-wise clean image and haze separately using a hazy image as a condition. A diffusion model was proposed in~\cite{Stevens2025} to reconstruct high-quality 3D ultrasound from sparsely sampled elevation planes, outperforming conventional and deep learning interpolators. Chen \textit{et al.} proposed a latent dynamic diffusion model to synthesize realistic ultrasound videos from static images, enhancing video classification performance and addressing the scarcity of labeled ultrasound video datasets~\cite{Chen2024}. Dominguez \textit{et al.} introduced a physics-based diffusion model tailored for ultrasound imaging, incorporating a custom scheduler that simulates acoustic wave propagation to generate more realistic synthetic data~\cite{Dominguez2024}.

Although effective, these methods depend on clean training data, which is often unavailable in ultrasound, and tend to be computationally complex. For example,  some diffusion-based methods can only perform offline processing even on powerful GPUs~\cite{10423849}, highlighting the need for faster methods~\cite{Zhu2025DiG}. Self-supervised models address the reliance on paired clean data by learning directly from noisy inputs without ground truth, which are discussed in the next section.

\subsection{Self-Supervised Methods}

Noise2Noise~\cite{Lehtinen2018} demonstrated that denoising networks can be trained using only pairs of independently corrupted images, eliminating the need for clean targets by leveraging the statistical independence of noise. This method showed that learning to map one noisy image to another could still lead to effective noise suppression. Building on this work, self-supervised denoising methods often rely on strategies like masking, random corruption, or predicting pixels based on their surroundings. For example, Noise2Void and Noise2Self~\cite{Krull2019, Batson2019} hide certain pixels and train the network to predict their values using neighboring pixels, assuming that nearby pixels share similar information while noise is random. Neighbor2Neighbor~\cite{Huang2021} extends this idea by identifying pairs of nearby pixels that are likely to share the same underlying structure and using them to train the network. 

However, many of these approaches were developed for natural or biomedical microscopy images, which often differ significantly from ultrasound data in terms of noise characteristics and image formation physics. Ultrasound images exhibit speckle, coherent artifacts, and depth-dependent attenuation, all of which challenge generic self-supervised assumptions. Thus, domain-specific adaptations are necessary for successful ultrasound denoising. Therefore, several self-supervised approaches have emerged. Zhang \textit{et al.}~\cite{ZHANG2021102018} aimed to reconstruct high-quality ultrasound plane-waves from raw channel data by training a self-supervised network in an inverse problem approach. Sharifzadeh~\textit{et al.}~\cite{10584319} introduced an aberration correction strategy, where the network was trained using pairs of images that shared structural content but differed in aberration. This approach allowed the model to learn to suppress aberration artifacts without requiring clean ground truth images by leveraging the variability in the aberrated inputs. A deblurring-based masked image modeling framework was proposed in \cite{Kang2024} to improve ultrasound image representation learning by leveraging its high noise-to-signal ratio, achieving state-of-the-art performance on classification and segmentation tasks.

One promising approach is Deep Coherence Learning~\cite{Cho2024}, which eliminates the need for paired noisy-clean training data by enforcing coherence constraints between multiple plane-waves, improving image quality while maintaining high frame rates. Despite its advantages, this method necessitates training data that closely matches the domain of the inference data, limiting its effectiveness under new imaging conditions, transducers, or anatomical structures that were not well-represented during training. Additionally, it requires multi-day GPU training, as reported in their study. 

Zero-shot denoising eliminates the need for pre-collected training datasets by leveraging the intrinsic properties of noisy images to perform self-consistent denoising. The ZS-N2N framework~\cite {Mansour2023} has demonstrated that noise removal can be performed without access to high-quality images, making it particularly suitable for applications where acquiring high-quality ground truth is infeasible. In this work, we adapt ZS-N2N to ultrasound plane-wave denoising; our approach considers low-angle compounded images as inherently noisy and enhances their quality by leveraging self-supervised learning. Without requiring external supervision, it efficiently suppresses noise, offering a robust and data-efficient denoising solution.

The rest of the paper is designed as follows. The proposed method is described in Section \ref{Method}. Then, the experimental setups, datasets, evaluation metrics, and training strategy are detailed in Section \ref{Experiments}, followed by the comparison results in~\ref{Results}. The  discussions are in Section \ref{Discussion}, and the paper is concluded in Section \ref{Conclusion}. 

\section{Methods}\label{Method}

\begin{figure*}[t]
  \centering
  \includegraphics[width=1\textwidth]{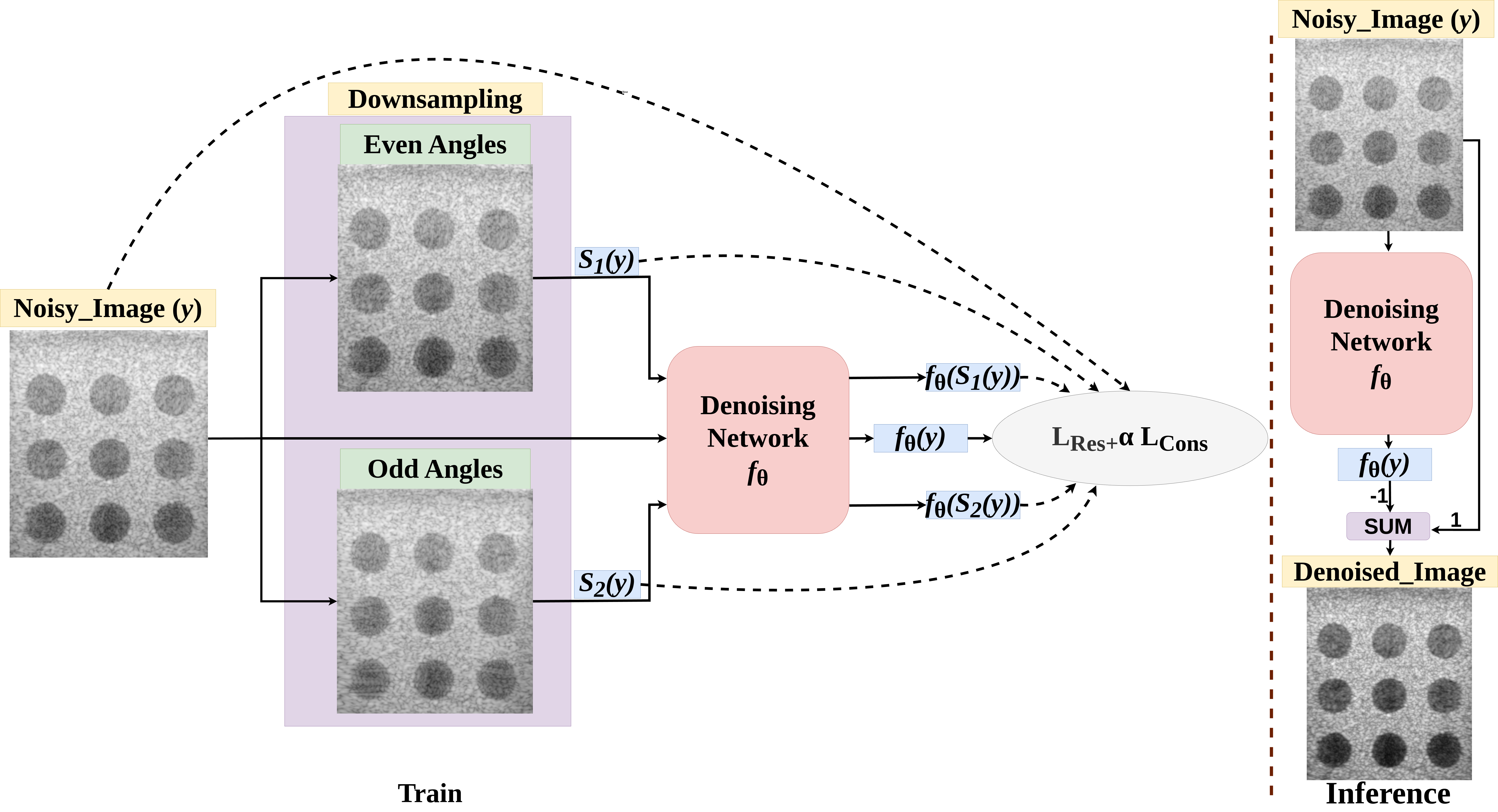}
  \caption{An overview of the proposed method. A low-compounding plane-wave image \( y \)  is decomposed into even-angle \( S_1(y) \) and odd-angle \( S_2(y) \) subsets. A lightweight network \( f_{\theta} \) is trained using a combined residual and consistency loss function (\( L_{\text{Res}} + \alpha L_{\text{Cons}} \)) to learn noise characteristics. Once trained, in the inference phase, the network estimates and removes the noise component from \( y \), producing a denoised output.}
  \label{Zero-shot-denoising-architecture}
\end{figure*}

We propose a zero-shot denoising framework for CPWC images by leveraging the angular diversity inherent in the acquisition process. \textcolor{black}{Specifically, a compounded image \( y \), formed from a low number of transmissions, is decomposed into two subsets: one constructed using odd-indexed angles and the other using even-indexed angles (Fig.~\ref{Zero-shot-denoising-architecture}). Each subset yields a compounded image with increased noise levels as a result of diminished angular compounding.}

\textcolor{black}{A lightweight CNN \( f_{\theta} \) is subsequently trained on this pair using a combination of symmetric residual and consistency loss functions. The training process encourages the network to learn incoherent components, primarily noise and artifacts, between the two inputs. Since the original image \( y \) is composed of both subsets, the trained network can estimate the noise residuals present in \( y \) and subtract them, resulting in an enhanced image with improved contrast and reduced noise.}

\subsubsection{Image Downsampling}
\textcolor{black}{CPWC consists of a set of different steering angles, denoted as \( \{P_1, P_2, \dots, P_k\} \), where \( k \) represents the number of transmissions over a range of \(-N^\circ \text{ to } +N^\circ \). To construct training pairs, we define two sampling functions, \( S_1(\cdot) \) and \( S_2(\cdot) \), which extract disjoint subsets of the available angles:}

\begin{equation}
S_1(y) = \{ P_i \mid i \in \mathcal{I}_1 \}, \quad S_2(y) = \{ P_j \mid j \in \mathcal{I}_2 \},
\end{equation}
where \( \mathcal{I}_1 \) and \( \mathcal{I}_2 \) are non-overlapping index sets (e.g., odd and even indices). The two resulting compounded images, \( S_1(y) \) and \( S_2(y) \), serve as inputs to the network. Although each is individually noisy, they share a common anatomical structure while differing in artifacts. This enables the model to learn angle-dependent artifacts directly from the input without requiring any external supervision or clean reference data.

\subsubsection{Plane-Wave Denoising}
Given a plane-wave image \( y \), the proposed approach involves training a lightweight network \( f_{\theta} \) to estimate the discrepancies between two sampled subsets. By learning the relationship between \( S_1(y) \) and \( S_2(y) \), the network disentangles correlated anatomical structures from the artifacts, thereby enhancing the original noisy plane-wave image. A baseline loss function for this mapping is as follows:

\begin{equation}
L = \left\| f_{\theta} (S_1(y)) - S_2(y) \right\|_1,
\end{equation}
where $\left\|.\right\|_1$ denotes \( L_1 \) norm. Once trained, the network can be directly applied to denoise the original compounded image using \( x = f_{\theta}(y) \), where \( x \) denotes the denoised output. However, experimental evidence suggests that residual learning provides more effective noise suppression \cite{Zhang2017}. Instead of directly predicting the clean image, the network is trained to estimate the noise residual and subtract it from the input. The corresponding residual formulation is:
\begin{equation}
L = \left\| S_1(y) - f_{\theta}(S_1(y)) - S_2(y) \right\|_1.
\end{equation}
Inspired by symmetric training strategies in self-supervised representation learning \cite{Chen2021}, we adopt a symmetric residual loss that enforces bidirectional consistency between the two downsampled images:

\begin{align}
L_{\text{Res.}} = \frac{1}{2} \Big( 
&\left\| S_1(y) - f_{\theta}(S_1(y)) - S_2(y) \right\|_1 \nonumber \\
+\, &\left\| S_2(y) - f_{\theta}(S_2(y)) - S_1(y) \right\|_1 
\Big).
\end{align}

\subsubsection{Consistency Loss}

To ensure that the denoised image remains structurally faithful to the original domain and to mitigate over-smoothing, we introduce a gradient-based consistency loss function. This loss encourages the gradient of the denoised image to match the gradient of the original input image \( y \), helping to preserve important edge and texture information. 

However, directly using \( \nabla y \) as a reference may propagate high-frequency noise into the loss, as the gradient operator amplifies noise. To address this, we replace \( \nabla y \) with its smoothed version, obtained by convolving \(y\) with a Gaussian kernel \( G_\sigma \) followed by the gradient operator. The consistency loss is then defined as:

\begin{equation}
L_{\text{Cons.}} = \left\| \nabla \left( y - f_{\theta}(y) \right) -  \nabla (G_\sigma *y) \right\|_1,
\end{equation}

where \( * \) denotes convolution and \( G_\sigma \) is a Gaussian filter.

\subsubsection{Total Loss Function}
The final training loss is expressed as follows:
\begin{equation}
L_{\text{Total}} = L_{\text{Res.}} + \alpha \cdot L_{\text{Cons.}},
\end{equation}
where \( \alpha \) is the consistency loss coefficient. We train the proposed method using \( L_{\text{Total}} \). After training, the network is used to denoise the original noisy plane-wave image by computing:
\begin{equation}
x = y - f_{\theta}(y),
\end{equation}
where \( x \) denotes the final denoised image.

\section{Experiments}\label{Experiments}
This section begins by describing the datasets used for evaluation. Subsequently, the implementation details are outlined, followed by a description of the evaluation metrics employed to quantify performance.
\subsection{Datasets}
We evaluate the proposed method on a diverse range of datasets comprising simulation, phantom, and \textit{in vivo} acquisitions. This collection is specifically curated to support standardized assessment of ultrasound beamforming and denoising algorithms, ensuring both fair comparisons and reproducibility across future studies.
\subsubsection{Simulation Contrast}
Image containing anechoic cysts (from PICMUS dataset~\cite{Liebgott}) arranged horizontally and vertically within a fully developed speckle background, specifically designed to evaluate contrast performance.
\subsubsection{Experimental Phantom}
\textcolor{black}{The experimental phantom data were acquired using a Verasonics Vantage 256 research ultrasound scanner equipped with an L11-5v linear array transducer (Verasonics Inc., Redmond, WA). The phantom used in the study was a CIRS Multi-Purpose Ultrasound Phantom (Model 040GSE, CIRS, Norfolk, VA), which contains both wire targets and anechoic cysts embedded in a tissue-mimicking speckle background. The transducer parameters were as follows: element pitch of 0.30~mm, element width of 0.27~mm, element height of 5~mm, elevation focus at 20~mm, and a total of 128 active elements, resulting in an effective aperture of 38.4~mm. The transmit frequency was 5.208~MHz, the sampling frequency was 20.832~MHz, the pulse bandwidth was 67\%, and the excitation comprised 2.5 cycles. A total of 75 steered plane-waves were transmitted over an angular span of $-16^{\circ}$ to $+16^{\circ}$. The acquired data were stored in both RF (modulated) and IQ (demodulated) formats and organized in HDF5 files for post-processing.}
\subsubsection{\textit{in vivo}}
The image of the cross-sectional (denoted as Carotid Cross-sectional, CC) and longitudinal (denoted as Carotid Longitudinal, CL) views of the carotid artery from a healthy volunteer (from PICMUS dataset~\cite{Liebgott}). 
\subsection{Implementation Details}

The proposed framework was implemented in PyTorch and trained on log-compressed B-mode images reconstructed using Delay-and-Sum beamforming. We used a lightweight convolutional architecture consisting of only two $3 \times 3$ convolutional layers with 48 channels, each followed by a Leaky ReLU activation function (negative slope = 0.2), and a final convolution $1 \times 1$ for refinement of the characteristics. The total parameter count is under 22k, making the network highly efficient. Training was performed using stochastic gradient descent (SGD) with an initial learning rate of $0.001$. The model was optimized over 1000 iterations for each test image in a zero-shot manner using a single NVIDIA RTX 4090 GPU. The consistency loss coefficient $\alpha$ was heuristically set to $0.25$, and $G_\sigma$ is a $3 \times 3$ Gaussian filter. \textcolor{black}{All experiments were conducted using five steering angle compounding ($k=5$ and $N=16$) as the original input. The angular spectrum was set to range from \(-16^\circ\) to \(16^\circ\) in our experiments (with \(N = 16\)), which was divided into 75 discrete angles, including \(-16^\circ\) and \(16^\circ\). To select the 5 angles, we sampled these 75 angles with the same interval of \(\frac{75}{5}\), resulting in the angles \([-16^\circ, -8^\circ, 0^\circ, 8^\circ, 16^\circ]\), respectively. For subset selection, we then classified the angles based on their index parity: the even indexes \([-16^\circ, 0^\circ, 16^\circ]\) were grouped into one set, and the remaining angles \([-8^\circ, 8^\circ]\) were placed in the other set.}

\subsection{Evaluation Metrics}\label{Evaluation metrics}
We quantitatively evaluated the performance of our method using three metrics: Contrast-to-Noise Ratio (CNR), generalized Contrast-to-Noise Ratio (gCNR)~\cite{Rodriguez}, and Kolmogorov–Smirnov (KS) test.
\subsubsection{Contrast-to-Noise Ratio (CNR)}  
CNR quantifies the separation between the Region Of Interest (ROI) and the background by measuring the ratio of their mean intensity difference to the combined noise level. It is computed as:
\begin{align}
\mathrm{CNR} = 20 \log_{10}\left(\frac{\left| \mu_{\text{ROI}} - \mu_B \right|}{\sqrt{\frac{\sigma^2_{\text{ROI}} + \sigma^2_B}{2}}}\right),
\label{eq23}
\end{align}
where $\mu$ and $\sigma$ represent the mean and standard deviation of the ROI and background regions, respectively.

\subsubsection{Generalized Contrast-to-Noise Ratio (gCNR)}  
As CNR can be sensitive to dynamic range scaling~\cite{Rodriguez}, gCNR was introduced to address this limitation. It is defined as:
\begin{align}
\mathrm{gCNR} = 1 - \int_{-\infty}^{\infty} \min\{p_{\text{ROI}}(x), p_B(x)\} \, dx,
\label{eq24}
\end{align}
where $p_{\text{ROI}}$ and $p_B$ are the intensity histograms of the ROI and background. gCNR measures the degree of histogram overlap between regions, with values closer to 1 indicating a clearer separation.

\subsubsection{Kolmogorov-Smirnov (KS) Test}  
To evaluate speckle preservation, we use the KS test to compare the cumulative distribution functions of pixel intensities between the processed and original images. \textcolor{black}{Since our model operates directly on log-compressed data, the speckle distribution in this domain does not strictly follow a Rayleigh model. Therefore, rather than testing against a theoretical distribution, we performed a two-sample KS test between the denoised output and the reference 5-angle compounded image within fully developed speckle regions. Passing the KS test (i.e., failing to reject the null hypothesis with $p > 0.05$) indicates that the statistical properties of the speckle are preserved and that no artificial texture has been introduced.}

\section{Results}\label{Results}

 \textcolor{black}{We compare our proposed method with classical, supervised, and zero-shot denoising approaches. As a classical baseline, we include BM3D~\cite{Gan2015}. For supervised baselines, we evaluate MSLAE~\cite{Perdios} and DPWDPM~\cite{Asgariandehkordi2024}, which were originally developed for beamformed RF data. To have a fair comparison, both models were trained on both beamformed RF and B-mode images from the same simulation dataset~\cite{Asgariandehkordi2024}. To further benchmark our zero-shot framework, we included zero-shot self-supervised methods such as ZS-N2N~\cite{Mansour2023}, Noise2Self (N2S)~\cite{Batson2019}, and Self2Self (S2S)~\cite{Ko2023}. Quantitative evaluations are conducted using CNR, gCNR, and the KS test, as described in Section~\ref{Evaluation metrics}.} All visual results are in B-mode, shown using a dynamic range of $-80$ dB, normalized to the $[0, 1]$ range for consistency. In all of the quantitative and visual results, 5 CPWC stands for compounding with 5 steering angles, and 75 CPWC stands for compounding with 75 steering angles.

\subsubsection{Simulation Results}

\begin{figure*}[t]
  \centering
  \includegraphics[width=\textwidth]{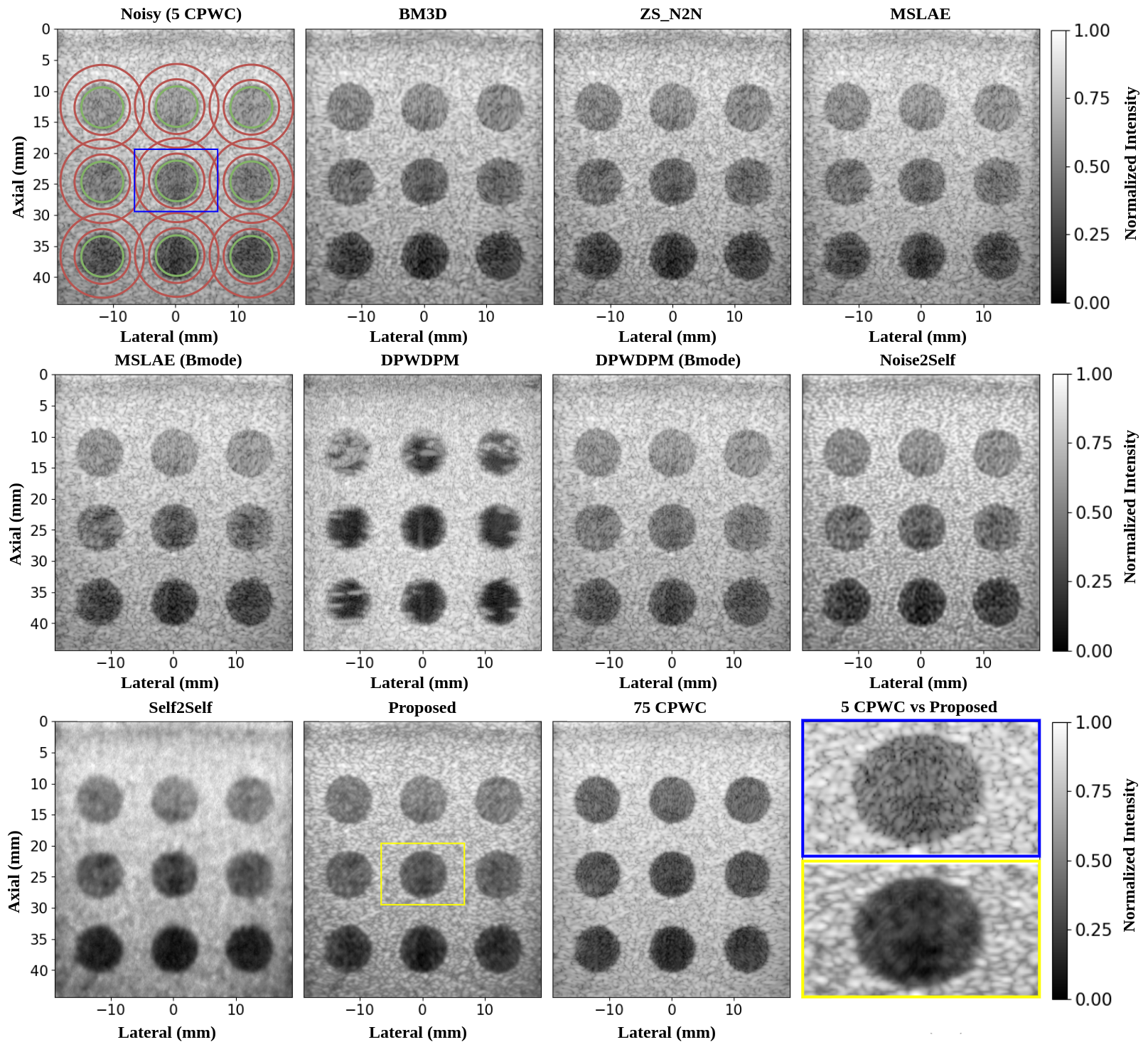}
  \caption{Visual comparison of the simulation results. All images are displayed with a dynamic range of $-80$~dB normalized to $[0, 1]$. The green and red regions indicate the ROI and the background areas, respectively. The zoomed-in images correspond to the blue and yellow rectangles in the noisy image (5 CPWC) and the proposed method, respectively.}
  \label{SIMULATION_ZERO_SHOT_DENOISING}
\end{figure*}

\textcolor{black}{To evaluate denoising performance under controlled conditions, simulation cyst data was used. The quantitative contrast evaluation was performed using gCNR and CNR metrics. For each cyst, 20 distinct foreground windows were selected to compute CNR values, following the statistical protocol described in Section~\ref{Evaluation metrics}. Mean and variance values were calculated across three cyst rows to ensure a robust and unbiased estimation. The cyst ROIs and background areas used for these measurements are shown in Fig.~\ref{SIMULATION_ZERO_SHOT_DENOISING}. sing green and red circles, respectively.}

\textcolor{black}{Fig.~\ref{SIMULATION_ZERO_SHOT_DENOISING} and Table~\ref{TAB1_SIMULATION_ZEROSHOT_DENOISING} present visual and quantitative comparisons of the proposed method against classical (BM3D), supervised (MSLAE, DPWDPM), and self-supervised or zero-shot (ZS\_N2N, N2S, S2S) approaches. For fairness, MSLAE and DPWDPM were retrained directly on B-mode data to avoid domain shift from their original RF-based configurations.}

\textcolor{black}{The proposed method consistently achieves the highest contrast across all cyst rows, with gCNR values of $0.82$, $0.81$, and $0.80$ and corresponding CNR values of $3.9$~dB, $3.7$~dB, and $3.6$~dB. These results surpass both supervised and self-supervised baselines, including N2S ($0.68$–$0.69$ gCNR) and S2S ($0.76$–$0.78$ gCNR). The retrained supervised models (MSLAE-Bmode and DPWDPM-Bmode) show moderate improvements compared to their original RF-trained versions but still exhibit residual blurring or weaker contrast recovery. All methods pass the KS test, confirming statistical consistency with the speckle characteristics of the fully compounded reference.}

Fig.~\ref{SIMULATION_ZERO_SHOT_DENOISING} provides qualitative comparisons. Our method enhances contrast and preserves fine structural details while avoiding the blurring or artifacts present in competing techniques. A zoomed-in view shows the visual gain achieved compared to the noisy input (5 CPWC), bringing our result closer to the high-quality 75 CPWC reference.

As for the runtime, for the image in Fig.~\ref{SIMULATION_ZERO_SHOT_DENOISING} with a size 300 $\times$ 384, our method takes 6 seconds (which includes both training and inference). For the same image, BM3D and ZS-N2N take 1 and 3 seconds, respectively. N2S and S2S also take more than 10 seconds. DPWDPM and MSLAE, which are applied to RF image of size 1200 $\times$ 384, take 1.2 and 0.2 seconds (for inference only), respectively. Our method is not currently optimized for speed; in the~\ref{Discussion} section (Discussion section), we will outline strategies to substantially accelerate it.

\begin{table}[h]
\caption{Quantitative results on simulation data. The contrast metrics are averaged across each row of cysts. All of the methods passed the KS test.}
\centering
\resizebox{\columnwidth}{!}{
\begin{tabular}{|l|c|c|c|}
    \hline
    \textbf{Method} & \textbf{Row} & \textbf{gCNR ($\uparrow$)} & \textbf{CNR (dB) ($\uparrow$)} \\
    \hline
    \multirow{3}{*}{Noisy Image} 
        & 1 & $0.56 \pm 0.011$ & $2.6 \pm 0.41$  \\
        & 2 & $0.59 \pm 0.011$ & $2.5 \pm 0.57$  \\
        & 3 & $0.58 \pm 0.014$ & $2.4 \pm 0.15$ \\
    \hline

    \multirow{3}{*}{BM3D} 
        & 1 & $0.63 \pm 0.007$ & $3.04 \pm 0.01$  \\
        & 2 & $0.64 \pm 0.005$ & $2.9 \pm 0.11$  \\
        & 3 & $0.63 \pm 0.005$ & $2.8 \pm 0.08$  \\
    \hline

    \multirow{3}{*}{ZS\_N2N} 
        & 1 & $0.56 \pm 0.012$ & $2.7 \pm 0.06$  \\
        & 2 & $0.58 \pm 0.010$ & $2.6 \pm 0.09$  \\
        & 3 & $0.58 \pm 0.009$ & $2.5 \pm 0.60$  \\
    \hline

    \multirow{3}{*}{MSLAE} 
        & 1 & $0.57 \pm 0.003$ & $2.7 \pm 0.01$  \\
        & 2 & $0.60 \pm 0.006$ & $2.6 \pm 0.168$  \\
        & 3 & $0.58 \pm 0.006$ & $2.6 \pm 0.12$  \\
    \hline

    \multirow{3}{*}{MSLAE (Bmode)} 
        & 1 & $0.56 \pm 0.007$ & $2.6 \pm 0.01$ \\
        & 2 & $0.59 \pm 0.006$ & $2.5 \pm 0.11$  \\
        & 3 & $0.57 \pm 0.006$ & $2.5 \pm 0.08$  \\
    \hline

    \multirow{3}{*}{DPWDPM} 
        & 1 & $0.77 \pm 0.010$ & $3.5 \pm 0.42$  \\
        & 2 & $0.77 \pm 0.006$ & $3.4 \pm 0.44$  \\
        & 3 & $0.78 \pm 0.005$ & $3.4 \pm 0.65$  \\
    \hline

    \multirow{3}{*}{DPWDPM (Bmode)} 
        & 1 & $0.56 \pm 0.012$ & $2.7 \pm 0.05$  \\
        & 2 & $0.59 \pm 0.010$ & $2.5 \pm 0.10$  \\
        & 3 & $0.58 \pm 0.010$ & $2.5 \pm 0.24$  \\
    \hline

    \multirow{3}{*}{N2S} 
        & 1 & $0.68 \pm 0.010$ & $2.9 \pm 0.12$ \\
        & 2 & $0.69 \pm 0.010$ & $2.8 \pm 0.17$  \\
        & 3 & $0.68 \pm 0.020$ & $2.7 \pm 0.10$  \\
    \hline

    \multirow{3}{*}{S2S} 
        & 1 & $0.76 \pm 0.010$ & $3.0 \pm 0.07$  \\
        & 2 & $0.77 \pm 0.004$ & $3.2 \pm 0.28$  \\
        & 3 & $0.78 \pm 0.003$ & $3.1 \pm 0.20$  \\
    \hline

    \multirow{3}{*}{Proposed} 
        & 1 & $0.82 \pm 0.003$ & $3.9 \pm 0.01$  \\
        & 2 & $0.81 \pm 0.004$ & $3.7 \pm 0.30$  \\
        & 3 & $0.80 \pm 0.003$ & $3.6 \pm 0.23$  \\
    \hline

    \multirow{3}{*}{75 CPWC} 
        & 1 & $0.89 \pm 0.001$ & $5.8 \pm 0.10$  \\
        & 2 & $0.86 \pm 0.002$ & $5.6 \pm 0.30$  \\
        & 3 & $0.86 \pm 0.002$ & $5.4 \pm 0.23$  \\
    \hline
\end{tabular}
}
\label{TAB1_SIMULATION_ZEROSHOT_DENOISING}
\end{table}

\subsubsection{Phantom Results}

\begin{table}[h]
\caption{Quantitative results on phantom data for various denoising methods. Contrast metrics (GCNR and CNR) are reported for each cyst. All outputs passed  the KS test.}
\centering
\resizebox{\columnwidth}{!}{
\begin{tabular}{|l|c|c|c|}
    \hline
    \textbf{Method} & \textbf{Cyst} & \textbf{gCNR ($\uparrow$)} & \textbf{CNR (dB) ($\uparrow$)} \\
    \hline
    \multirow{2}{*}{Noisy Image (5 angles)} 
        & 1 & $0.73 \pm 0.002$ & $2.3 \pm 0.30$ \\
        & 2 & $0.65 \pm 0.002$ & $2.2 \pm 0.22$  \\
    \hline
    
    \multirow{2}{*}{BM3D} 
        & 1 & $0.79 \pm 0.005$ & $2.7 \pm 0.17$ \\
        & 2 & $0.73 \pm 0.001$ & $2.6 \pm 0.15$  \\
    \hline

    \multirow{2}{*}{ZS\_N2N} 
        & 1 & $0.75 \pm 0.005$ & $2.4 \pm 0.13$ \\
        & 2 & $0.70 \pm 0.003$ & $2.3 \pm 0.14$  \\
    \hline

    \multirow{2}{*}{MSLAE} 
        & 1 & $0.73 \pm 0.003$ & $2.4 \pm 0.12$  \\
        & 2 & $0.73 \pm 0.002$ & $2.2 \pm 0.17$  \\
    \hline

    \multirow{2}{*}{MSLAE (Bmode)} 
        & 1 & $0.73 \pm 0.005$ & $2.4 \pm 0.15$  \\
        & 2 & $0.63 \pm 0.001$ & $2.5 \pm 0.10$  \\
    \hline

    \multirow{2}{*}{DPWDPM} 
        & 1 & $0.74 \pm 0.004$ & $2.5 \pm 0.16$  \\
        & 2 & $0.72 \pm 0.003$ & $2.8 \pm 0.22$  \\
    \hline

    \multirow{2}{*}{DPWDPM (Bmode)} 
        & 1 & $0.73 \pm 0.006$ & $2.3 \pm 0.12$  \\
        & 2 & $0.65 \pm 0.003$ & $2.2 \pm 0.14$  \\
    \hline

    \multirow{2}{*}{N2S} 
        & 1 & $0.81 \pm 0.001$ & $2.5 \pm 0.14$  \\
        & 2 & $0.75 \pm 0.001$ & $2.3 \pm 0.09$  \\
    \hline

    \multirow{2}{*}{S2S} 
        & 1 & $0.84 \pm 0.003$ & $2.9 \pm 0.24$ \\
        & 2 & $0.80 \pm 0.003$ & $2.8 \pm 0.16$  \\
    \hline

    \multirow{2}{*}{Proposed} 
        & 1 & $0.88 \pm 0.002$ & $3.5 \pm 0.53$  \\
        & 2 & $0.84 \pm 0.001$ & $3.1 \pm 0.19$  \\
    \hline

    \multirow{2}{*}{75 CPWC} 
        & 1 & $0.93 \pm 0.006$ & $4.7 \pm 0.50$  \\
        & 2 & $0.83 \pm 0.004$ & $3.2 \pm 0.30$  \\
    \hline
\end{tabular}
}
\label{TAB1_PHANTOM_ZEROSHOT_DENOISING}
\end{table}

\begin{figure*}[t]
  \centering
  \includegraphics[width=1\textwidth]{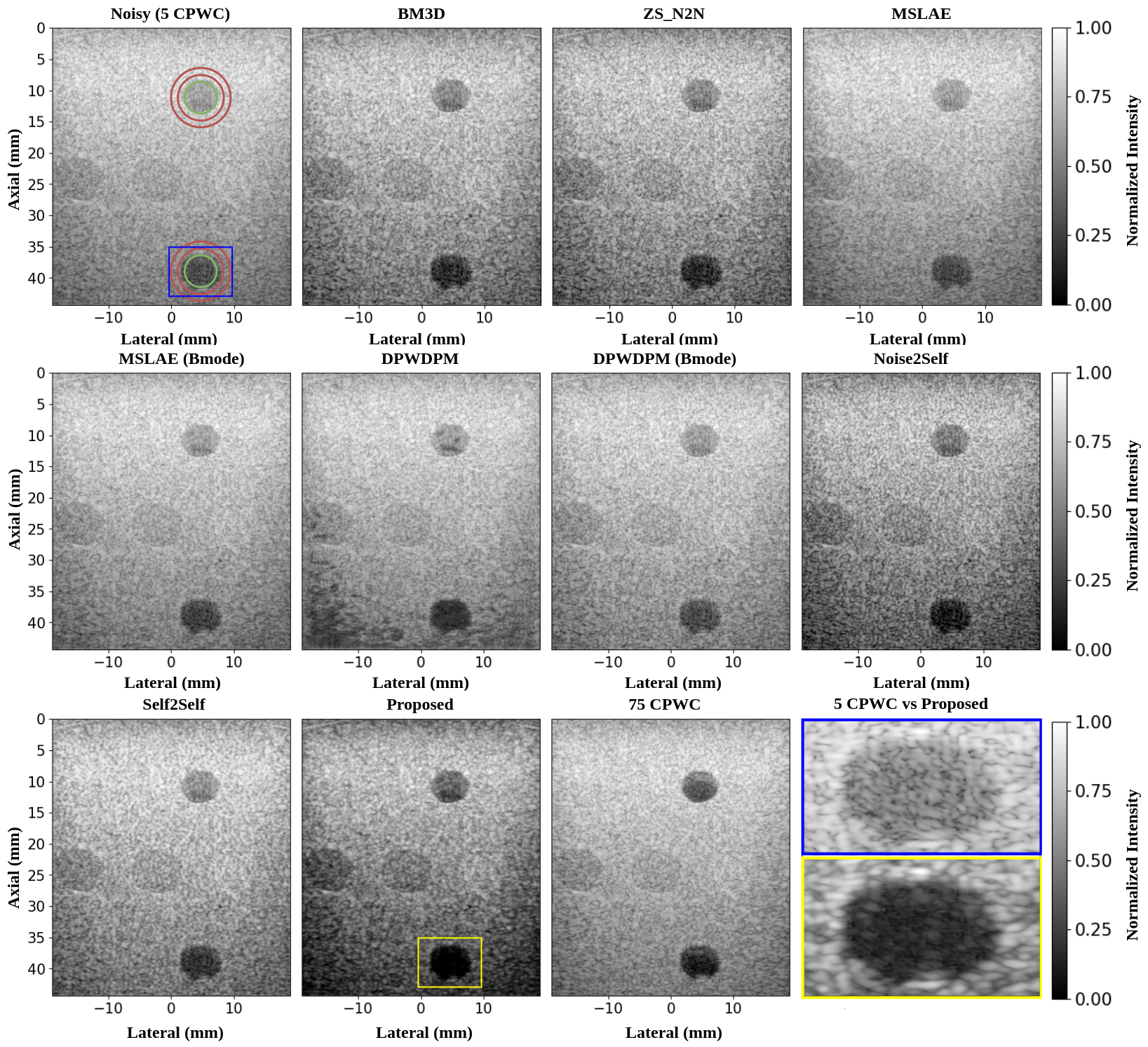}
  \caption{Visual comparison on the phantom data. All images are displayed with a dynamic range of $-80$~dB normalized to $[0, 1]$. The green and red regions indicate the ROI and the background areas, respectively. The zoomed images correspond to the blue and yellow rectangles in the noisy image (5 CPWC) and the proposed method, respectively.}
  \label{PHANTOM_ZERO_SHOT_DENOISING}
\end{figure*}

\textcolor{black}{The phantom dataset consists of two anechoic cysts embedded in a tissue-mimicking background. Cyst~1 is located near the transducer, while Cyst~2 lies at a greater depth. ROIs for the cysts and the background were defined using circular masks, as shown in Fig.~\ref{PHANTOM_ZERO_SHOT_DENOISING} (green for foreground and red for background). Quantitative contrast evaluation was performed using gCNR and CNR metrics computed from these ROIs.}

\textcolor{black}{Quantitative contrast metrics are summarized in Table~\ref{TAB1_PHANTOM_ZEROSHOT_DENOISING}.For the quantitative evaluation, CNR metrics of each cyst were computed using a statistical approach, where 10 distinct foreground windows were selected. The 75~CPWC image serves as the high-quality reference, with Cyst~1 yielding gCNR~(0.93) and CNR~(4.7~dB), and Cyst~2 reaching gCNR~(0.83) and CNR~(3.2~dB). These values establish the baseline for comparison.}

\textcolor{black}{Among the classical and supervised baselines, BM3D and the retrained MSLAE and DPWDPM models provide moderate denoising performance. BM3D achieves gCNR~(0.79) and CNR~(2.7~dB) for Cyst~1 but fails to significantly enhance deeper regions. The retrained B-mode versions of MSLAE and DPWDPM offer slightly improved homogeneity (e.g., DPWDPM-Bmode: gCNR~(0.73), CNR~(2.3~dB)), yet both exhibit mild texture loss and oversmoothing at greater depths.}

\textcolor{black}{The zero-shot methods—including ZS\_N2N, N2S, and S2S perform better at preserving cyst boundaries while providing varying degrees of noise suppression. ZS\_N2N achieves gCNR~(0.75) and CNR~(2.4~dB) for Cyst~1, but offers limited visual improvement. N2S improves contrast to gCNR~(0.81) and CNR~(2.5~dB), while S2S further enhances results with gCNR~(0.84) and CNR~(2.9~dB), though some residual speckle persists in homogeneous areas.}

\textcolor{black}{The proposed zero-shot framework demonstrates the highest overall and most consistent performance across both cysts. For Cyst~1, it achieves gCNR~(0.88) and CNR~(3.5~dB), and for Cyst~2, gCNR~(0.84) and CNR~(3.1~dB), surpassing all other methods and closely matching the 75~CPWC reference. All methods passed the KS test, indicating that the speckle pattern was preserved post-denoising. Visual comparisons in Fig.~\ref{PHANTOM_ZERO_SHOT_DENOISING} show that the proposed method yields sharper cyst boundaries and more homogeneous speckle, providing a strong balance between denoising strength, contrast enhancement, and texture fidelity.}

\subsubsection{\textit{in vivo} results}

To assess the clinical utility of our method, we evaluated it on two \textit{in vivo} scans: CC and CL. These images capture vascular anatomy embedded in tissue with moderate to low contrast, representative of real-world diagnostic conditions. We report both quantitative and qualitative comparisons with classical, supervised, and zero-shot baselines.

\paragraph{CC Dataset}

\begin{figure*}[t]
  \centering
  \includegraphics[width=1\textwidth]{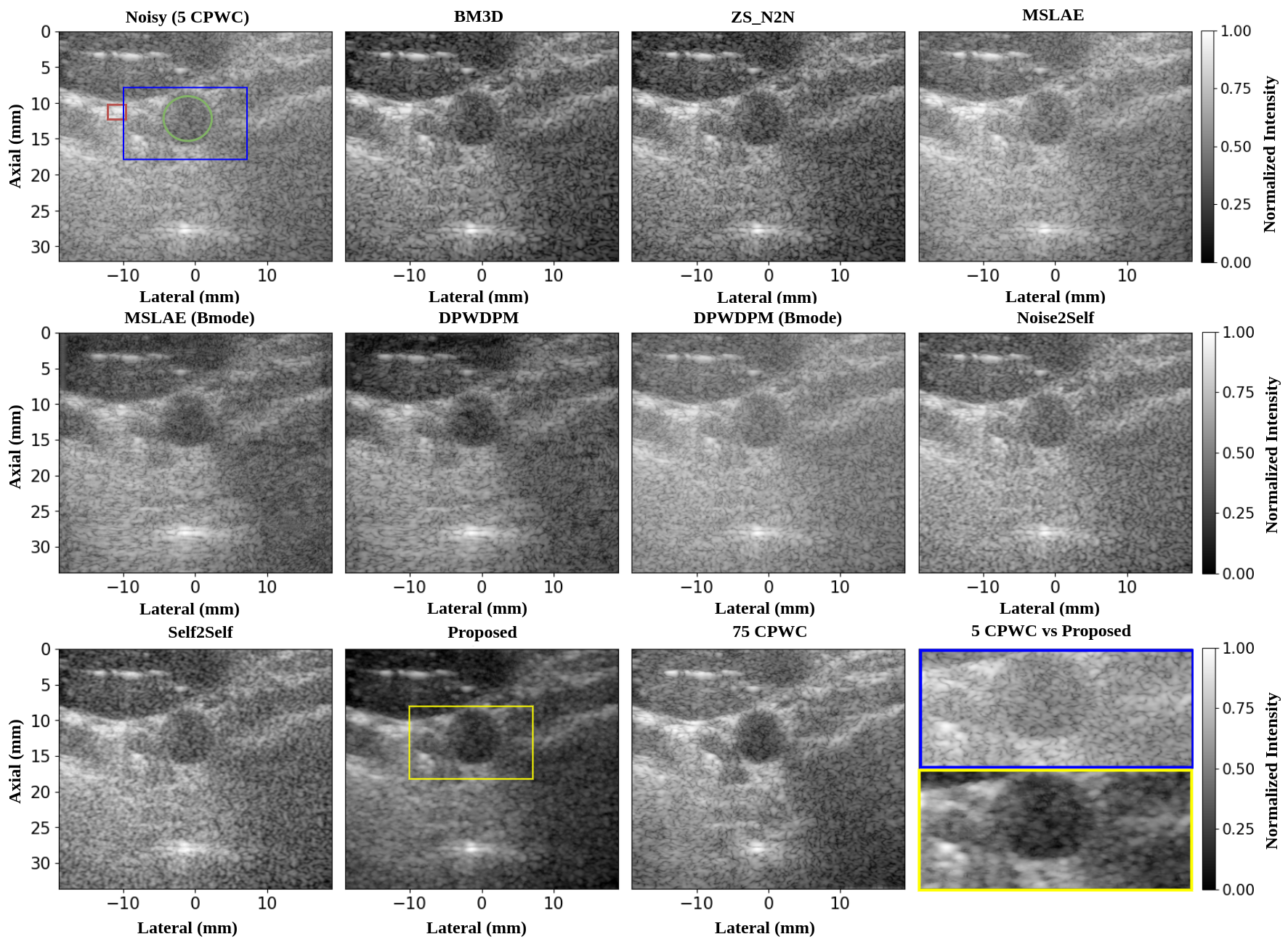}
  \caption{Visual comparison of \textit{in vivo} results (CC data). All images are displayed with a dynamic range of $-80$~dB normalized to $[0, 1]$. The green and red regions indicate the ROI and the background areas, respectively. The zoomed images correspond to the blue and yellow rectangles in the noisy image (5 CPWC) and the proposed method, respectively.}
  \label{IN-VIVO2_ZEROSHOT_DENOISING}
\end{figure*}

\textcolor{black}{Fig.~\ref{IN-VIVO2_ZEROSHOT_DENOISING} and Table~\ref{TAB1_IN-VIVO2_ZEROSHOT_DENOISING} summarize the comparison between the proposed zero-shot denoising framework and other reference methods, including BM3D, ZS\_N2N, MSLAE, DPWDPM, N2S, and S2S, on \textit{in vivo} cross-sectional data. The 75 CPWC image serves as the high-quality reference, yielding gCNR (0.76) and CNR (1.2~dB), which establish the baseline for comparison. For the quantitative evaluation, the contrast metrics were calculated using a statistical procedure in which 10 separate foreground regions were selected. Overall, the proposed method achieves the highest performance with gCNR (0.78) and CNR (1.3~dB), outperforming both classical and self-supervised baselines.}

\textcolor{black}{Among the zero-shot methods, BM3D and ZS\_N2N moderately improve image contrast but tend to smooth fine tissue boundaries. N2S and S2S further enhance contrast (up to gCNR 0.72 and CNR 1.01~dB) but retain residual noise in homogeneous areas. The supervised approaches, MSLAE and DPWDPM, even after retraining on B-mode data, exhibit limited improvement due to domain shift effects. MSLAE (B-mode) achieves gCNR (0.69) and CNR (0.80~dB), while DPWDPM (B-mode) reaches gCNR (0.65) and CNR (0.90~dB), with mild over-smoothing observed in both.}

\textcolor{black}{Visually, the proposed zero-shot method effectively suppresses noise while preserving tissue structure and boundary sharpness, producing results that closely resemble the 75 CPWC reference. As shown in Fig.~\ref{IN-VIVO2_ZEROSHOT_DENOISING}, it maintains homogeneous backgrounds and enhances contrast without sacrificing speckle realism. All methods passed the KS test.}

\begin{table}[h]
\caption{Quantitative results on \textit{In vivo} cross-sectional data for different denoising methods. All outputs passed  the KS test. The ROI and the background regions are shown in Fig.~\ref{IN-VIVO2_ZEROSHOT_DENOISING}. All of the methods passed the KS test.}
\centering
\resizebox{\columnwidth}{!}{
\begin{tabular}{|l|c|c|}
    \hline
    \textbf{Method} & \textbf{gCNR ($\uparrow$)} & \textbf{CNR (dB) ($\uparrow$)}  \\
    \hline
    Noisy Image (5 angles)   & $0.65 \pm 0.008$ & $0.83 \pm 0.15$ \\
    \hline
    BM3D                     & $0.69 \pm 0.007$ & $1.01 \pm 0.43$  \\
    \hline
    ZS\_N2N                  & $0.73 \pm 0.010$ & $0.91 \pm 0.35$  \\
    \hline
    MSLAE                    & $0.67 \pm 0.006$ & $0.84 \pm 0.11$  \\
    \hline
    MSLAE (Bmode)            & $0.69 \pm 0.016$ & $0.80 \pm 0.10$\\
    \hline
    DPWDPM                   & $0.70 \pm 0.020$ & $0.83 \pm 0.20$ \\
    \hline
    DPWDPM (Bmode)           & $0.65 \pm 0.013$ & $0.90 \pm 0.25$ \\
    \hline
    N2S                      & $0.67 \pm 0.008$ & $0.86 \pm 0.27$  \\
    \hline
    S2S                & $0.72 \pm 0.007$ & $1.01 \pm 0.32$ \\
    \hline
    Proposed & $0.78 \pm 0.006$ & $1.30 \pm 0.28$ \\
    \hline
    75 CPWC                  & $0.76 \pm 0.006$ & $1.20 \pm 0.20$ \\
    \hline
\end{tabular}
}
\label{TAB1_IN-VIVO2_ZEROSHOT_DENOISING}
\end{table}

\paragraph{CL Dataset}

\begin{figure*}[t]
  \centering
  \includegraphics[width=1\textwidth]{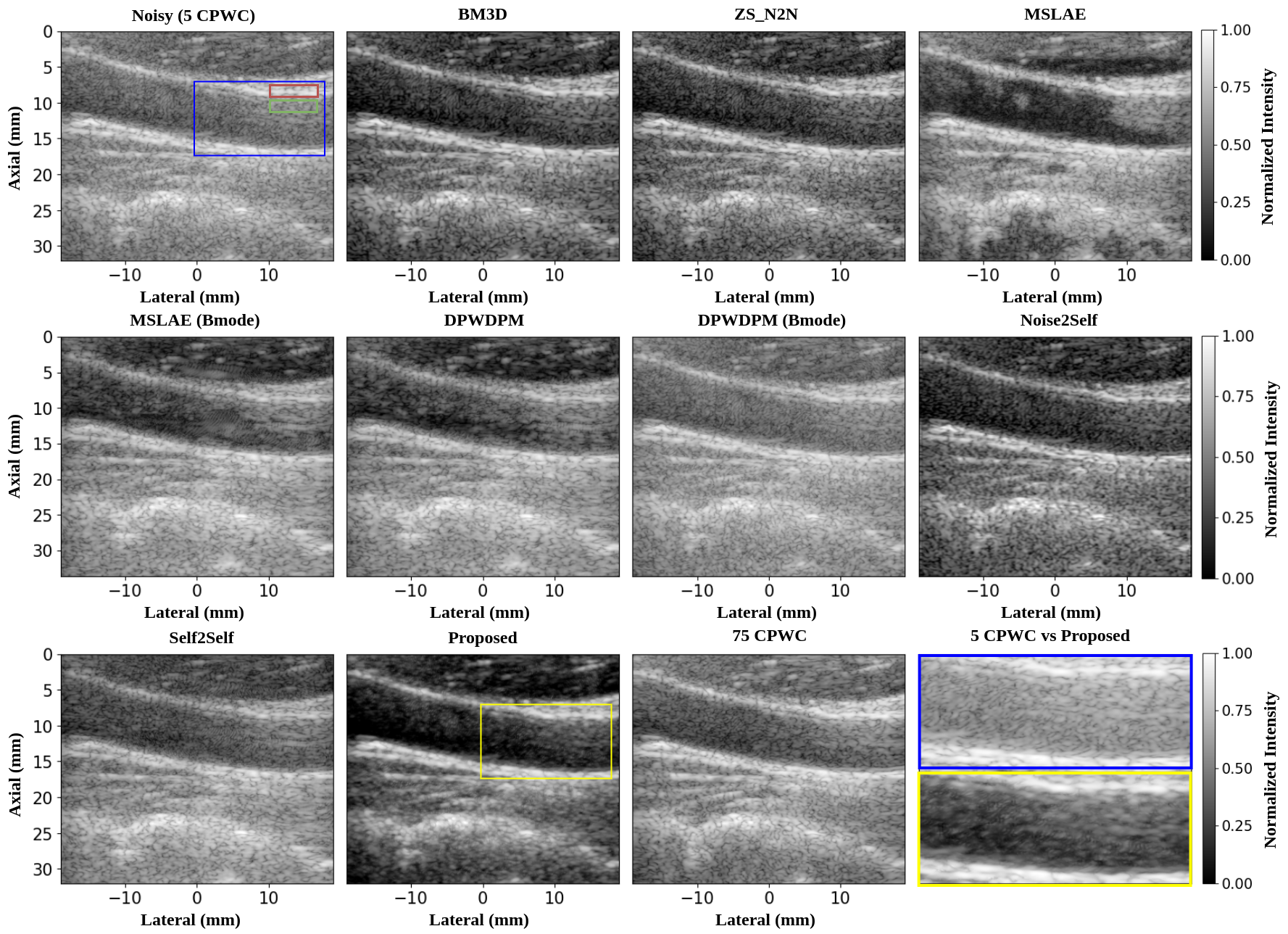}
  \caption{Visual comparison of \textit{in vivo} results (CL data). All images are displayed with a dynamic range of $-80$~dB normalized to $[0, 1]$. The green and red regions indicate the ROI and the background areas, respectively. The zoomed images correspond to the blue and yellow rectangles in the noisy image (5 CPWC) and the proposed method, respectively.}
  \label{IN-VIVO_ZEROSHOT_DENOISING}
\end{figure*}

 \begin{table}[h]
\caption{Quantitative results on \textit{in vivo} longitudinal data for different denoising methods. All outputs passed the KS test.}
\centering
\resizebox{\columnwidth}{!}{
\begin{tabular}{|l|c|c|}
    \hline
    \textbf{Method} & \textbf{gCNR ($\uparrow$)} & \textbf{CNR (dB) ($\uparrow$)} \\
    \hline
    Noisy Image (5 angles)   & $0.65 \pm 0.006$ & $2.4 \pm 0.20$ \\
    \hline
    BM3D                     & $0.75 \pm 0.007$ & $3.1 \pm 0.16$  \\
    \hline
    ZS\_N2N                  & $0.75 \pm 0.005$ & $3.0 \pm 0.24$ \\
    \hline
    MSLAE                    & $0.67 \pm 0.016$ & $2.7 \pm 0.30$  \\
    \hline
    MSLAE (Bmode)            & $0.61 \pm 0.016$ & $2.2 \pm 0.25$  \\
    \hline
    DPWDPM                   & $0.65 \pm 0.006$ & $2.2 \pm 0.16$  \\
    \hline
    DPWDPM (Bmode)           & $0.61 \pm 0.016$ & $2.2 \pm 0.27$  \\
    \hline
    N2S                      & $0.76 \pm 0.006$ & $2.6 \pm 0.10$  \\
    \hline
    S2S                & $0.67 \pm 0.003$ & $2.4 \pm 0.30$  \\
    \hline
    Proposed & $0.88 \pm 0.006$ & $3.6 \pm 0.20$  \\
    \hline
    75 CPWC                  & $0.82 \pm 0.060$ & $3.2 \pm 0.20$  \\
    \hline
\end{tabular}
}
\label{TAB1_IN-VIVO_ZEROSHOT_DENOISING}
\end{table}

\textcolor{black}{A comprehensive comparison of denoising outcomes on longitudinal \textit{in vivo} data is shown in Fig.~\ref{IN-VIVO_ZEROSHOT_DENOISING} and Table~\ref{TAB1_IN-VIVO_ZEROSHOT_DENOISING}. The evaluation spans classical, supervised, and self-supervised approaches, benchmarked against the proposed zero-shot method. To obtain consistent quantitative measures, CNR values for each cyst were computed using a statistical sampling strategy with 10 distinct foreground windows. The 75 CPWC compounded image was used as the high-quality reference (gCNR = 0.82, CNR = 3.2~dB). Among all methods, the proposed framework achieved the most favorable balance between contrast enhancement and speckle fidelity, reaching gCNR = 0.88 and CNR = 3.6~dB, which surpasses both learning-based and traditional counterparts.}\\
\textcolor{black}{In terms of individual performance, methods such as BM3D and ZS\_N2N showed moderate improvements in contrast ($\text{gCNR} \approx 0.75$, $\text{CNR} \approx 3.0~\text{dB}$), but these came at the expense of fine structural detail. Self-supervised alternatives, including N2S and S2S, achieved slightly sharper boundaries but left traces of residual noise, limiting their quantitative gains. The supervised models, MSLAE and DPWDPM, along with their retrained B-mode variants, struggled to adapt to the \textit{in vivo} data domain, displaying lower contrast metrics (e.g., MSLAE (Bmode): $\text{gCNR} = 0.61$, $\text{CNR} = 2.2~\text{dB}$). These differences highlight the advantage of the proposed zero-shot method, which avoids domain mismatch by training directly on the test image itself.
Qualitative results in Fig.~\ref{IN-VIVO_ZEROSHOT_DENOISING} further support these findings. The proposed method retains clear vessel wall boundaries and anatomical detail, while the other methods suffer from artifacts.}

\textcolor{black}{Overall, the proposed approach demonstrates the best trade-off between noise suppression, contrast enhancement, and speckle preservation on \textit{in vivo} imaging.}

\section{Discussion}\label{Discussion}
\textcolor{black}{This study introduces a zero-shot self-supervised denoising method tailored for plane-wave ultrasound imaging. Unlike most existing deep learning approaches that require either paired datasets or large-scale training data, our method leverages a \textcolor{black}{physics-aware} angle-based self-supervision strategy. This allows it to \emph{learn directly from the noisy test image} without the need for training data.}

\textcolor{black}{Quantitative and qualitative results across simulation, phantom, and \textit{in vivo} datasets demonstrate the robustness and generalizability of the proposed zero-shot denoising framework. In simulation experiments, the method consistently outperforms baselines that do not require a dedicated training dataset, such as BM3D, ZS-N2N, S2S, and N2S in both gCNR and CNR metrics across all cyst depths. Compared to supervised models such as MSLAE and DPWDPM, which were trained on synthetic datasets, the proposed approach achieves superior contrast without introducing artifacts, particularly in deeper regions where conventional methods encounter limitations due to lower signal to noise ratio.}

\textcolor{black}{In phantom experiments, which introduce real-world attenuation and speckle characteristics, the proposed approach again delivers strong performance. While BM3D performs reasonably well in shallow cysts, it tends to oversmooth deeper structures. In contrast, our method achieves the highest gCNR (0.84) and CNR (3.1) among all methods except the 75-angle compounded reference. Notably, these results are achieved without any retraining or adaptation to phantom data, highlighting the method’s resilience to domain shifts.}

\textcolor{black}{In \textit{in vivo} studies, the framework maintains superior contrast and anatomical fidelity in both shallow and deep vascular targets. On the CL scan, the proposed method surpasses all baselines in both gCNR (0.88) and CNR (3.6), exceeding even the 75-angle reference. On the more challenging CC scan, it also yields the highest gCNR (0.78) and matches the top CNR (1.3), again without requiring any external training data. Visual inspections confirm sharper vessel boundaries and reduced background noise, particularly in low-contrast regions.}

\textcolor{black}{Another strength of our approach lies in its compact architecture. With fewer than 22k parameters and only two convolutional layers, the model maintains a lightweight structure that minimizes the chance of overfitting.}
\textcolor{black}{Although the proposed method is not yet optimized for real-time denoising, future work will focus on enabling online deployment. To achieve this, we plan to pretrain the model on a diverse collection of plane-wave datasets, allowing it to learn a robust initialization. With such a universal prior, the model would require only a few fine-tuning iterations on new data rather than the thousand of iterations it currently needs, substantially reducing the computational complexity.}

\textcolor{black}{The method's ability to generalize across simulated, phantom, and \textit{in vivo} domains highlights the potential for broader applicability across ultrasound imaging techniques such as focused imaging and synthetic aperture imaging. Extending the model to support various probe geometries, such as curved or phased arrays, is an area of future work.}

\textcolor{black}{Finally, while our method demonstrates stable and competitive results across different conditions, future studies should investigate its performance on larger and more diverse clinical datasets. Furthermore, evaluating its impact on downstream clinical tasks, such as elastography, localization microscopy, or segmentation, would provide additional validation and insights into its practical utility.}

\section{Conclusions}\label{Conclusion}
\textcolor{black}{In this work, we presented a zero-shot denoising framework tailored for plane-wave ultrasound imaging, leveraging paired subsets of compounded data to self-supervise a lightweight residual learning network. Our method requires no clean references or training datasets, making it practical for real-world deployment in clinical ultrasound systems. Through extensive experiments on simulation, experimental phantom, and \textit{in vivo} datasets, we demonstrated that the proposed approach effectively reduces noise while preserving structural details and anatomical fidelity. The consistency and smoothing strategies further enhance robustness across varying imaging conditions. Despite its simplicity and low parameter count, our method competes favorably with supervised deep learning and classical denoising techniques, and in many cases, surpasses them. These results establish our approach as a promising, efficient solution for high-quality ultrasound reconstruction, particularly in settings with limited acquisition angles or a lack of training data.}

\bibliographystyle{IEEEtran}

\end{document}